\documentclass[journal=aaembp,manuscript=article]{achemso}

\usepackage{chemformula} 
\usepackage[T1]{fontenc} 
\usepackage{graphicx, xcolor} 
\usepackage{amsmath, gensymb, amssymb} 
\usepackage{hhline}
\usepackage[caption=false]{subfig}
\usepackage{siunitx}
\usepackage{miller}
\usepackage{xr}
\usepackage{upgreek}
\graphicspath{{Figures//}}
\mciteErrorOnUnknownfalse

\makeatletter
\newcommand*{\addFileDependency}[1]{
  \typeout{(#1)}
  \@addtofilelist{#1}
  \IfFileExists{#1}{}{\typeout{No file #1.}}
}
\makeatother

\newcommand*{\myexternaldocument}[1]{%
    \externaldocument{#1}%
    \addFileDependency{#1.tex}%
    \addFileDependency{#1.aux}%
}

\myexternaldocument{Supplementary}

\author{Suzanne Lancaster}
\email{suzanne.lancaster@namlab.com}
\affiliation[NaMLab]
{NaMLab gGmbH, N\"{o}thnitzer Str. 64a, 01187 Dresden, Germany}
\author{Maximilien Remillieux}
\affiliation[INL]
{\'{E}cole Sup\'{e}rieure de Chimie Physique 
\'{E}lectronique de Lyon, 3 Rue Victor Grignard, 69100 Villeurbanne, France}
\alsoaffiliation[NaMLab]
{NaMLab gGmbH, N\"{o}thnitzer Str. 64a, 01187 Dresden, Germany}
\author{Moritz Engl}
\affiliation[NaMLab]
{NaMLab gGmbH, N\"{o}thnitzer Str. 64a, 01187 Dresden, Germany}
\author{Viktor Havel}
\affiliation[NaMLab]
{NaMLab gGmbH, N\"{o}thnitzer Str. 64a, 01187 Dresden, Germany}
\author{Cl\'{a}udia Silva}
\affiliation[NaMLab]
{NaMLab gGmbH, N\"{o}thnitzer Str. 64a, 01187 Dresden, Germany}
\author{Xuetao Wang}
\affiliation[NaMLab]
{NaMLab gGmbH, N\"{o}thnitzer Str. 64a, 01187 Dresden, Germany}
\author{Thomas Mikolajick}
\affiliation[NaMLab]
{NaMLab gGmbH, N\"{o}thnitzer Str. 64a, 01187 Dresden, Germany}
\alsoaffiliation[IHM]{Institute of Semiconductors and Microsystems, Technische Universit\"{a}t Dresden, N\"{o}thnitzer Str 64., 01187 Dresden, Germany}
\author{Stefan Slesazeck}
\affiliation[NaMLab]
{NaMLab gGmbH, N\"{o}thnitzer Str. 64a, 01187 Dresden, Germany}

\title{Weight update in ferroelectric memristors with identical and non-identical pulses}

\begin{document}




\begin{abstract}
Ferroelectric tunnel junctions (FTJs) are a class of memristor which promise low-power, scalable, field-driven analog operation. In order to harness their full potential, operation with identical pulses is targeted. In this paper, several weight update schemes for FTJs are investigated, using either non-identical or identical pulses, and with time delays between the pulses ranging from 1\,$\mu$s to 10\,s. Experimentally, a method for achieving non-linear weight update with identical pulses at long programming delays is demonstrated by limiting the switching current \textit{via} a series resistor. Simulations show that this concept can be expanded to achieve weight update in a 1T1C cell by limiting the switching current through a transistor operating in sub-threshold or saturation mode. This leads to a maximum linearity in the weight update of 86\% for a dynamic range (maximum switched polarization) of 30 $\mu$C/cm$^2$. It is further demonstrated \textit{via} simulation that engineering the device to achieve a narrower switching peak increases the linearity in scaled devices to \textgreater 93 \% for the same range.
\end{abstract}

\section{Introduction}
 With the discovery of ferroelectricity in HfO$_2$ \cite{boscke2011ferroelectricity}, the possibility of integrating ferroelectric devices with CMOS circuitry became much more promising. Ferroelectric devices based on HfO$_2$ demonstrate fast switching speeds \cite{francois2022high, dahan2023sub}, power efficiency \cite{covi2021ferroelectric} and dense integration suitability \cite{de202228}. As such, they show great promise for applications such as neuromorphic \cite{ryu2019ferroelectric, gibertini2022ferroelectric} or edge computing \cite{o2018prospects}, and logic-in-memory \cite{reuben2023low, yang2020memory}. For these use cases, it is particularly interesting to consider multilevel memristive device operation, which enables synaptic operation \cite{begon2022scaled} and increases bit density \cite{goh2021high}. In particular, linear weight update schemes are desirable in order to improve accuracy in neuromorphic processors \cite{moon2019rram}. One of the main hindrances here is the application of switching pulses on-chip: in order to set a device into states of different levels, non-identical pulses are generally required \cite{oh2017hfzro}. 

 In the most ideal case, a ferroelectric device should update its weight when subjected to identical switching pulses. This would facilitate synaptic behaviour \cite{ielmini2019emerging} and simplify programming schemes in crossbar configurations \cite{gokmen2016acceleration}. Due to the switching of ferroelectrics, showing a field- and frequency-dependent distribution \cite{massarotto2022versatile}, weight update is classically unobtainable in this manner. A certain field will be applied across the ferroelectric on a given pulse, and all domains with a coercive field less than the applied field at the pulse frequency will switch, so that if backswitching is neglected, there will be no switchable domains when applying the next pulse. Polarization switching in hafnia ferroelectrics is understood to be governed by nucleation-limited switching \cite{tagantsev2002non}. In this model, when a switching pulse is applied, regions of the ferroelectric will switch. The time needed to switch a region is dominated only by the nucleation time, i.e. the domain expansion time is negligble. Each region can be described by a characteristic waiting time related to the nucleation rate in that region. For a longer switching time, the required voltage for domain nucleation is lower; conversely, for a higher voltage, domain nucleation will happen within a shorter pulse width. This leads to the well-known time-voltage tradeoff in ferroelectrics \cite{mulaosmanovic2020interplay} and indicates that increasing the voltage or pulse width is necessary for the nucleation of additional domains in multilevel programming schemes. In practice, diverse observations in the literature suggest that weight update with identical pulses can be achieved through extrinsic, rather than intrinsic effects \cite{lancaster2022multi, siannas2023electronic, vecchi2024evaluation}. 
 
 In Ferroelectric Field Effect Transistors (FeFETs), an accumulative weight update effect has been observed with identical pulses \cite{mulaosmanovic2020investigation}, while later experiments on write disturb indicated that this is not intrinsic to ferroelectric hafnia but may be stack-dependent \cite{hoffmann2022write}. Similarly in Ferroelectric Tunnel Junctions (FTJs), it was shown that weight update can be achieved with identical pulses at a time delay between switching pulses of 1$\mu$s \cite{lancaster2022multi}, but weight update greatly diminishes with increasing time delay between identical pulses \cite{siannas2023electronic}. The stronger weight update at short time delays is proposed to originate from extrinsic effects such as interfacial charge trapping \cite{vecchi2024evaluation}, residual fields when switching with fast pulses \cite{siannas2023electronic}, or ionic motion in metal-oxide electrodes. While it is possible to design circuitry to set multilevel states with non-identical pulses \cite{narayanan2022120db}, this adds additional constraints in terms of chip area, energy, and programming algorithms \cite{zhou2020application}. In that respect, multilevel operation using identical pulses is a much more favourable approach. 

 FTJs can be considered as memristors since the polarization modulates the device resistance and thus the current density at an applied read voltage \cite{ryu2019ferroelectric, chanthbouala2012solid, max2020hafnia}. In bilayer FTJs, as investigated here, the tunnelling through a ferroelectric/dielectric (MFDM) stack is modulated by the polarization state, which can be switched with an external field of positive (\textit{´On'}) or negative polarity (\textit{´Off'}). In the \textit{´On'} state, the polarization points towards the dielectric layer and when a read voltage applied on the MF electrode, tunnelling is mainly limited by the dielectric layer. This was originally assumed to lead to Fowler-Nordheim tunnelling through the stack \cite{max2020hafnia}, although recent simulations indicate that charge compensation leads instead to direct tunnelling through the dielectric layer \cite{lancaster2023reducing}. In the \textit{´Off'} state, both the ferroelectric and dielectric layers act as tunnel barriers. Partially switching the domains leads to states with an intermediate resistance \cite{max2020hafnia, begon2022scaled}. In this article, we will present different operating schemes for ferroelectric devices, using an FTJ as a model device. First, we will revisit switching with non-identical pulses as well as published approaches to switching with identical pulses and discuss their limitations, i.e. the necessity of strict timing conditions. Then, we will demonstrate a novel approach which allows weight update in ferroelectric devices using identical pulses on arbitrary time scales, when a resistor of a sufficiently high resistance is connected in series during the switching pulse. In this case, weight update is achieved with identical pulses, although it is non-linear, and the full memory window (MW) is limited by the RC constant of the system. A final scheme is proposed and demonstrated via simulations, where a weight update with identical pulses is performed while applying a current compliance to the switching circuit, which should allow a linear weight update spanning the maximum MW. These schemes greatly simplify the on-chip weight update of ferroelectric devices.

\section{Weight update with non-identical pulses}
\subsection{Characterization}
\begin{figure}[!hb]
\centering
		\includegraphics[height=5cm]{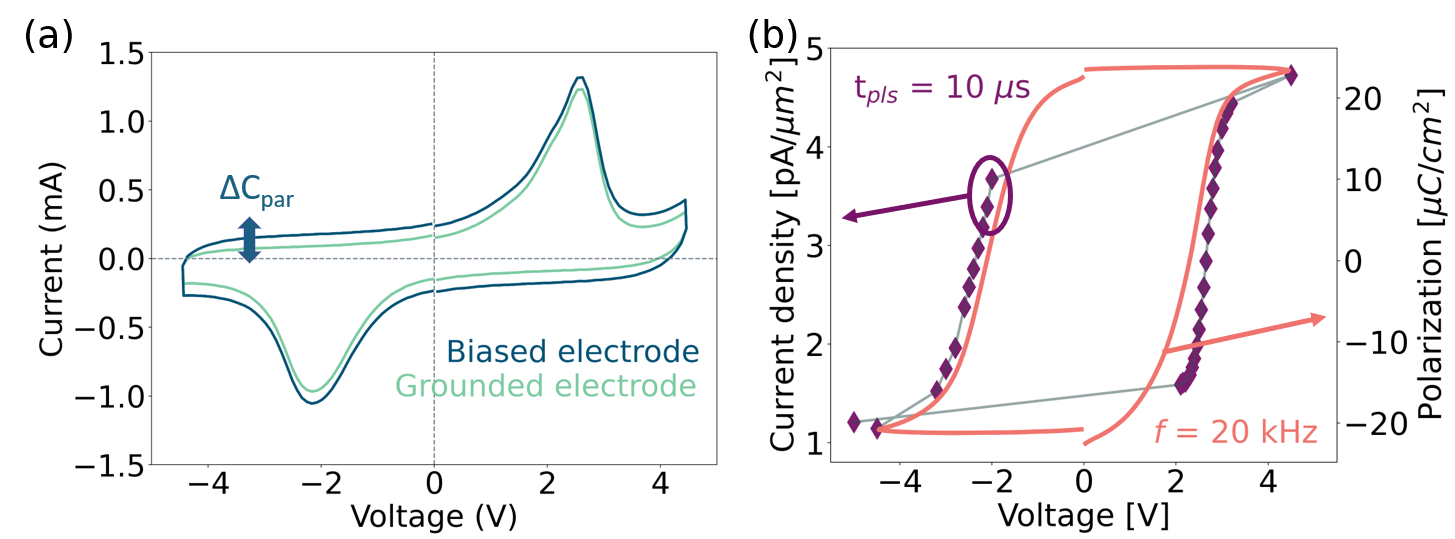}
	\captionsetup{font=footnotesize}
	\caption{(a) Current-voltage and (b) polarization-voltage curves (orange lines) for the 110\,nm diameter FTJ devices used in these experiments, measured at 20\,kHz. Purple diamonds represent the current density measured with pulses of 50\,ms, 1.3\,V, for pulses of 10\,$\mu$s at the voltages shown on the x-axis. The devices were woken-up at 100\,kHz, 4.5\,V, 1e$^4$ cycles.}
\label{Fig1_SwitchingBasic}
\end{figure}

 In these experiments, bilayer FTJs were used, with either 11\,nm ferroelectric Hf$_{0.5}$Zr${0.5}$O$_2$ (HZO) and a 1\,nm tunnel barrier (FTJ-1nm) to minimize switching and read voltages \cite{lancaster2023reducing} or 12\,nm HZO and a 2\,nm barrier \cite{max2020hafnia} (FTJ-2nm).  The full description of sample processing is given in the methods section. Figure \ref{Fig1_SwitchingBasic}a shows the I-V curve at 20\,kHz (after wake-up at 4.5\,V, 1e$^4$ cycles, 100\,kHz square pulses) curves measured during switching on an FTJ-1nm device. A higher background current is measured on the biased electrode due to an additional capacitive charging of the electrical setup, which needs to be considered when modelling the switching behaviour. Figure \ref{Fig1_SwitchingBasic}b shows the P-V loop extracted from (a) using a PUND technique (orange line, extracted from the I-V of the grounded electrode). The purple diamonds represent the FTJ current density measured with rectangular read pulses of 1.3\,V, 50\,ms after programming at the indicated voltages with 10\,$\mu$s programming pulses. It is clear that the FTJ read-out current follows the trend of the P-V loop, which indicates that the read-out current and polarization follow a linear relationship, as shown previously for FTJ-2nm devices \cite{lancaster2022multi}.

\subsection{Varying pulse width and time}
\begin{figure}[!ht]
\centering
		\includegraphics[height=10cm]{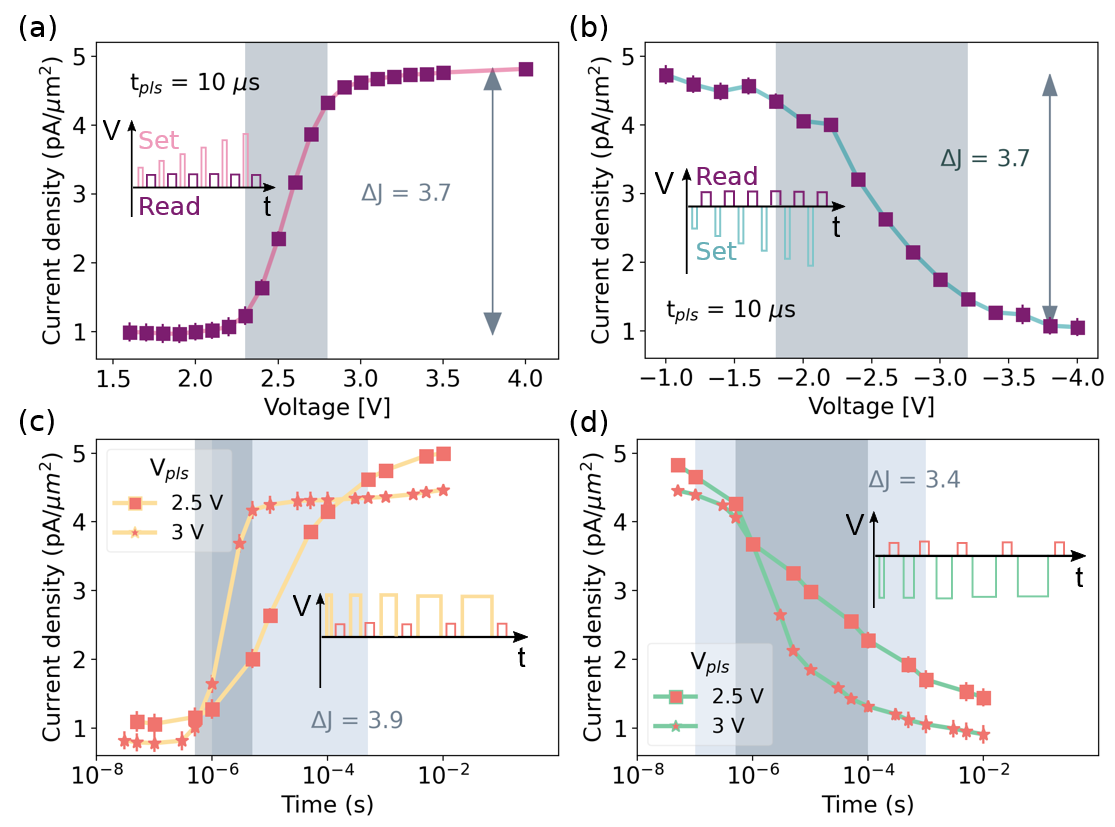}
	\captionsetup{font=footnotesize}
    \caption{(a), (b) Current density vs. applied voltage for pulsed weight update on an FTJ-1nm device using positive or negative pulses of increasing voltage magnitude and 10 $\mu$s pulse width. (c), (d) Current density vs. pulse time for pulsed weight update of an FTJ device using positive or negative pulses of increasing time width and fixed voltage amplitudes - 2.5\,V (squares) or 3\,V (stars). The FTJ state is read with a voltage pulse in between each applied set pulse, with the pulse trains indicated inset.}
\label{Fig2_SwitchingKinetics}
\end{figure}

The understanding of how a ferroelectric reacts to increasing or decreasing voltages or pulse widths is paramount to harnessing its synaptic function \cite{max2020hafnia}. A first approximation for appropriate partial switching pulses can be found from switching kinetics measurements, where the polarization switched under pulses of a given width and amplitude can be found. Then, one parameter (pulse width/pulse amplitude) can be fixed and the switching behaviour can be explored when varying the free parameter. Examples are plotted here for FTJ-1nm, either fixing the pulse width and increasing the voltage (fig \protect\ref{Fig2_SwitchingKinetics}a\&b) or fixing the voltage and increasing the pulse width (\protect\ref{Fig2_SwitchingKinetics}c\&d).

In addition, fig \protect\ref{Fig2_SwitchingKinetics}c\&d compare the switching behaviour at two fixed voltage amplitudes - a larger fixed voltage can be applied in order to reduce the pulse time needed to switch the ferroelectric, shifting the curve to smaller times. This voltage-time trade-off is inherent to hafnia ferroelectrics \cite{mulaosmanovic2020interplay} since they follow a nucleation-limited switching behaviour \cite{zhou2013wake} as has also been shown for FTJ devices \cite{lancaster2022multi}. The specific voltage-time relationship depends sensitively on material process parameters \cite{hsain2022role} and is therefore highly device-specific. FTJ devices in particular show an asymmetry in the weight update in positive and negative polarities, related to an internal field introduced by the dielectric layer \cite{lancaster2022investigating}, which modifies the applied bias in each switching direction and can be observed as an asymmetry in the switching I-V curves in positive and negative polarities (figure \ref{Fig1_SwitchingBasic}a). Despite a smaller negative coercive voltage, the voltage/time window for resetting the device is larger than for setting, due to a broader switching distribution in the negative polarity. If the switching amplitude is not large enough, it may be difficult to fully reset the device, even at long pulse times (see figure \protect\ref{Fig2_SwitchingKinetics}d).

\subsection{Generating an automata scheme}
\begin{figure}[!hb]
\centering
		\includegraphics[height=6cm]{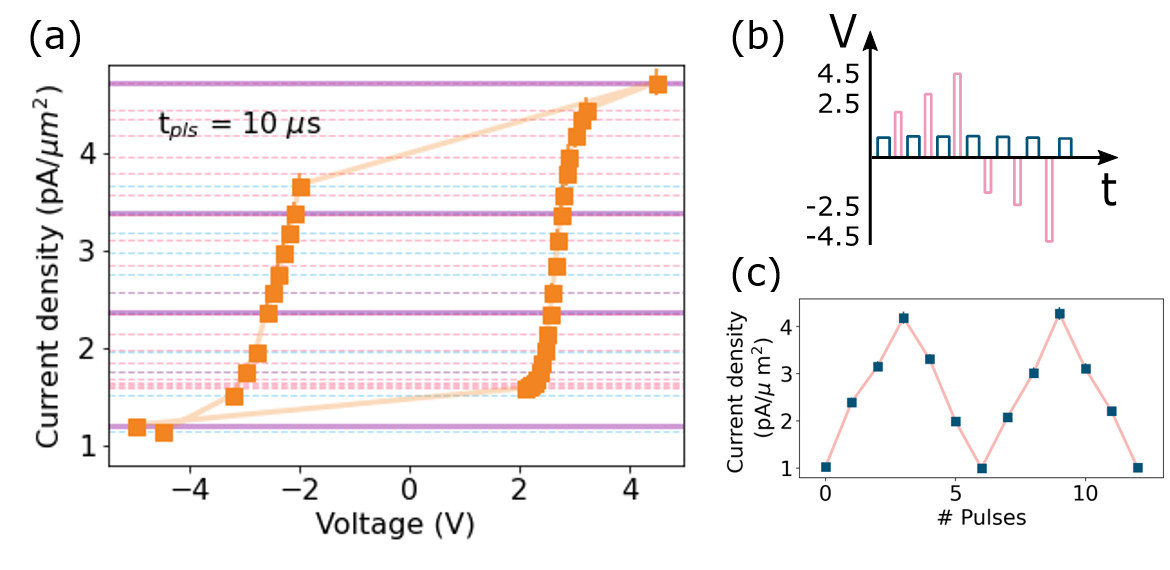}
	\captionsetup{font=footnotesize}
    \caption{(a) FTJ-1nm current densities measured for programming voltages with an interval of 0.1\,V, for positive and negative programming pulses. Red (blue) dashed lines indicate the current density reached with positive (negative) programming pulses, and purple lines indicate four states with a \textless 5\% overlap. (b) A programming scheme extracted from (a) for programming four states with asymmetric positive and negative voltages. (c) The corresponding device current densities measured when applying this programming scheme.}
\label{Fig3a_Automata}
\end{figure}
Based off of figure \ref{Fig2_SwitchingKinetics}, it is possible to produce pairs of pulses in positive and negative polarities which reach the same polarization state. This can be described in an \textit{automata} model which defines several states and the required pulse parameters to move between them. This can form the basis of a finite state machine and has applications in logic-in-memory circuits, where the low readout currents of the FTJ device can be exploited for low-power operation \cite{reuben2023low}. 

In order to generate an automata model based on a ferroelectric device, the experiments shown in \ref{Fig2_SwitchingKinetics}a\&b were first repeated with a voltage step of 0.1\,V. The results are plotted in \ref{Fig3a_Automata}a. The dashed horizontal lines represent the current densities achieved for voltage pulses with positive voltages (red) and negative voltages (blue). To define the automata, four equidistant states were targeted. States where positive and negative voltages give an overlapping current density (within $\pm 5\%$) were identified, and the most equidistant of those were chosen as the targeted states, highlighted with purple horizontal lines.

Figure \ref{Fig3a_Automata}b shows the programming scheme used to write the four states during programming; the corresponding current densities are plotted in figure \ref{Fig3a_Automata}c. The programming scheme could alternatively be written as a state machine. This defines four deterministic states with a separation of 1\,pA/$\mu$$m^2$ and the positive/negative voltages required to reach them. Here it should be noted that due to the ferroelectric switching mechanism, for intermediate states it is still necessary to know the current state of the device in order to know whether to apply a positive or negative switching pulse. During real device operation, this can be circumvented by first fully switching the device in one direction, and then partially switching back, in order to reach intermediate states without an initial read operation. Furthermore, as mentioned previously, the switching behaviour is highly device-specific and thus the required switching pulses will also depend sensitively on the device processing conditions.  

\section{Weight update with identical pulses}
\subsection{Time-correlated switching pulses}
\begin{figure}[!hb]
\centering
		\includegraphics[height=5cm]{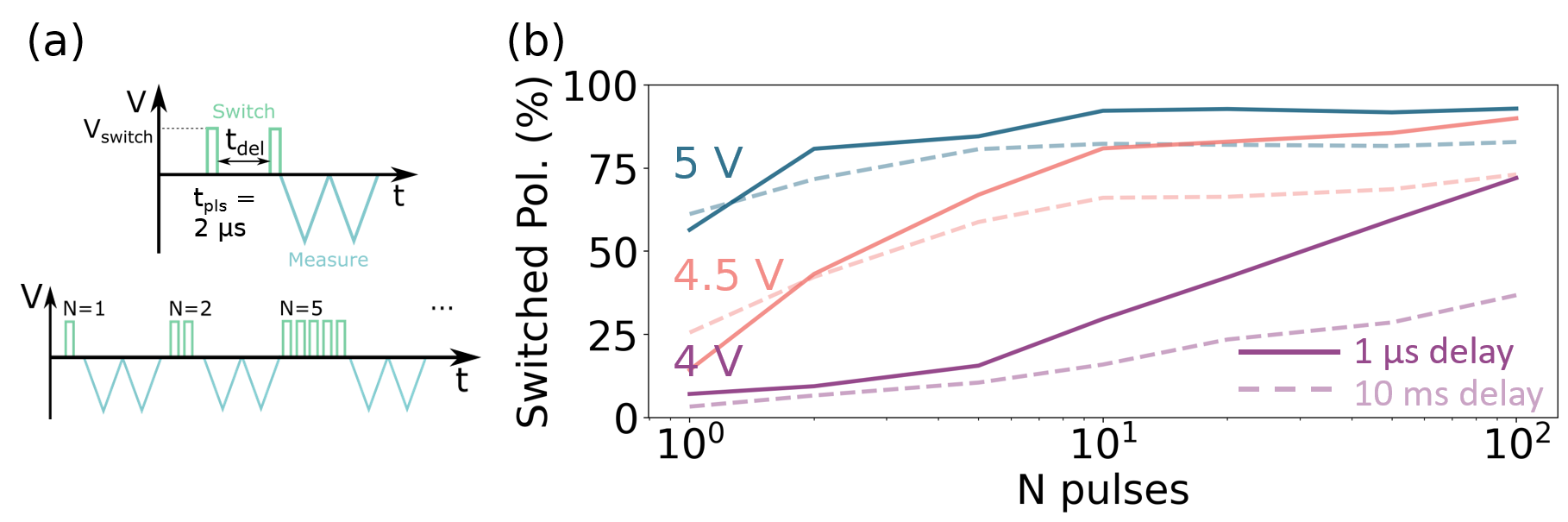}
	\captionsetup{font=footnotesize}
    \caption{(a)Pulse trains for measuring time-correlated switching with \textit{n} switching pulses. (b) Switched polarization measured for \textit{n} pulses of 10\,$\mu$s at different voltages and for a delay time between pulses of 1\,$\mu$s (solid lines) or 10\,ms (dashed lines), on FTJ-2nm.}
\label{Fig3_TimeCorr}
\end{figure}
The term ´time-correlated switching pulses' was coined by Siannas et al. \cite{siannas2023electronic} to describe the effect of a cumulative weight update behaviour when the time between switching pulses is kept very small. Such measurements have been demonstrated previously on our FTJ-2nm devices \cite{lancaster2022multi} and are revisited here. A schematic for this type of switching is shown in figure \ref{Fig3_TimeCorr}. The switching pulse width t$_{pls}$ is fixed at 2\,$\mu$s, and the switching pulse amplitude V$_{switch}$ is varied for different measurement series. N = [1, 100] pulses are applied with either a 1\,$\mu$s or 10\,ms delay (t$_{del}$) between pulses (where 10\,ms delays are chosen to represent switching on biological timescales of 100\,Hz \cite{jackson2013nanoscale}). The polarization is measured with an ND (Negative-Down) triangular pulse set; this value is then compared to the switching measured on a full switching pulse at the start of each measurement. The results are plotted in figure \ref{Fig3_TimeCorr}b, where solid (dashed) lines represent the polarization switched with a t$_{del}$ of 1\,$\mu$s (10\,ms).

To explain the effect of time-correlated pulses, it has previously been posited that the effective field in the capacitor does not drop to zero in very short delay times due to the capacitor discharging time \cite{siannas2023electronic}. It should be noted that ionic effects may also play a role for HZO on oxide electrodes, given the higher time constant of ionic motion \cite{halter2023multi}. However, the results shown here clearly reveal a continued switching effect for time delays far beyond the discharging time, which can be estimated from the device RC constant ($\sim$ 1$e^{-6}$\,s, based on the measured device capacitance \cite{benatti2023linking}). On longer timescales, the additional weight update can be attributed to a fluid imprint effect \cite{buragohain2019fluid} whereby injected carriers modify the probability of domain nucleation on timescales lower than the charge redistribution time \cite{vecchi2024evaluation}. If charges injected at the interface are particularly stable, this may eventually lead to the accumulative switching behaviour which has been observed \cite{mulaosmanovic2020investigation}.  However, such weight update will generally be strongly time-dependent and will lead to negligible weight update at time intervals of seconds or longer between pulses.

In summary, in the context of pulse-delay we can distinguish between three switching regimes and corresponding models. In the case of very short pulse delays, NLS-like switching kinetics which treat the pulse train as a continuous pulse can appropriately describe the weight update, where the integral of accumulated pulse duration (pulse width x number of pulses) is decisive for weight update, independently of the individual pulse width \cite{siannas2023electronic}. That is, the capacitors remain in a ‘excited’ state during the (< $\mu$s) pulse delay time. 

In contrast, for very long pulse delays (> s) this ‘excited’ state decays before the next pulse arrives (see figure \ref{Fig4_IdenticalResistor}a). Under such a condition pulses can be treated as independent and one can assume that all domains with a coercive field below the actual effective switching field (depending on pulse width according the NLS mode and non-ideal properties of the stack, e.g. charge trappingl) will be flipped. Thus, allowing the use of a Preisach model (time-independent model) to simulate the behaviour of the device. Any additional similar pulse would not cause additional weight update. 

Finally, there is a third regime at the transition between the two extremes. The separation between these three regimes in terms of pulse delay and its dependence on pulse width and pulse amplitude depends strongly not only on the ferroelectric material properties, but the full device stack (including interlayers and electrodes) and might strongly differ for the different device concepts (i.e. FeFET, FeCAP or FTJ). Moreover, endurance cycling causing wake-up and fatigue, or imprint during storage \cite{barbot2023dynamics} might further change the devices operation regime. Therefore, so far no satisfactory linear weight update using identical voltage pulses has been shown.

\subsection{Current-limited switching with resistor in series}
In order to achieve weight update with identical pulses and arbitrary time delays, current limiting schemes were applied. When a voltage pulse, V(t), is applied to a ferroelectric, the current, I(t) which flows has three components:
\begin{equation}
I(t) = A\frac{dP(t)}{dt} + C_{FE}\frac{dV_{FE}(t)}{dt} + I_{leak}(t)
\label{CurrentMainEq}
\end{equation}
where the first term corresponds to the switching current, proportional to the area of the capacitor, \textit{A} and the change in polarization \textit{$\frac{dP(t)}{dt}$}; the second term is the dielectric displacement current and is proportional to the ferroelectric capacitance, \textit{C$_{FE}$} and the voltage drop over the ferroelectric \textit{$\frac{dV_{FE}(t)}{dt}$}; and \textit{I$_{leak}$} is the leakage through the stack. 

During polarization switching, a current proportional to $\frac{dP(t)}{dt}$ will flow. The principle behind this type of weight update using identical pulses is to restrict the polarization switching charge \textit{Q$_{P}$} for a single pulse, by controlling both the polarization current flow \textit{I$_{P}$} through the device during switching as well as the respective switching time \textit{t$_{P}$}. In the ideal case the polarization charge update yields then \textit{Q$_{P}$ = I$_{P}$ * t$_{P}$}. The ferroelectric switching mechanism would remain field-driven, where the field across the ferroelectric device is self-limiting depending on the charge through the capacitor stack.

In the simplest case, the current control can be realized by a resistor, with a resistor much larger than the parasitic resistance of the setup and electrodes, i.e. $R_{ser} \gg R_{par}$ being inserted in series with the ferroelectric capacitor. Then, the effective voltage $V_{FTJ}$ that is applied to the FTJ is limited by the externally applied voltage $V_{app}$ and the resistance $R_{ser}$, according to: 
\begin{equation}
V_{FTJ} = V_{app} - R_{ser} * I_{fe}(t),
\label{EffectiveVoltage}
\end{equation}
where $I_{FTJ}(t)$, is the capacitor current according to equation \ref{CurrentMainEq} with the capacitance $C_{FE}$ replaced by $C_{FTJ}$. In that way, the field across the ferroelectric will no longer be determined by the externally applied field (minus a field drop across the dielectric), but rather will be self-limited by the polarization current flow through the stack and the external resistance \cite{massarotto2024analysis}. In the equation the $V_{FTJ}$ is considered to be the voltage drop over the whole device, i.e. the voltage drop over the ferroelectric will be lower \cite{park2021polarizing}. In this case the Preisach Model is extracted for the bilayer system which incorporates the influence of the additional dielectric layer \textit{via} a shift in the coercive field to higher values and a combined capacitance of the ferroelectric and dielectric $C_{FTJ}$ \cite{gibertini2022ferroelectric}.  

\begin{figure}[!hb]
\centering
		\includegraphics[height=10cm]{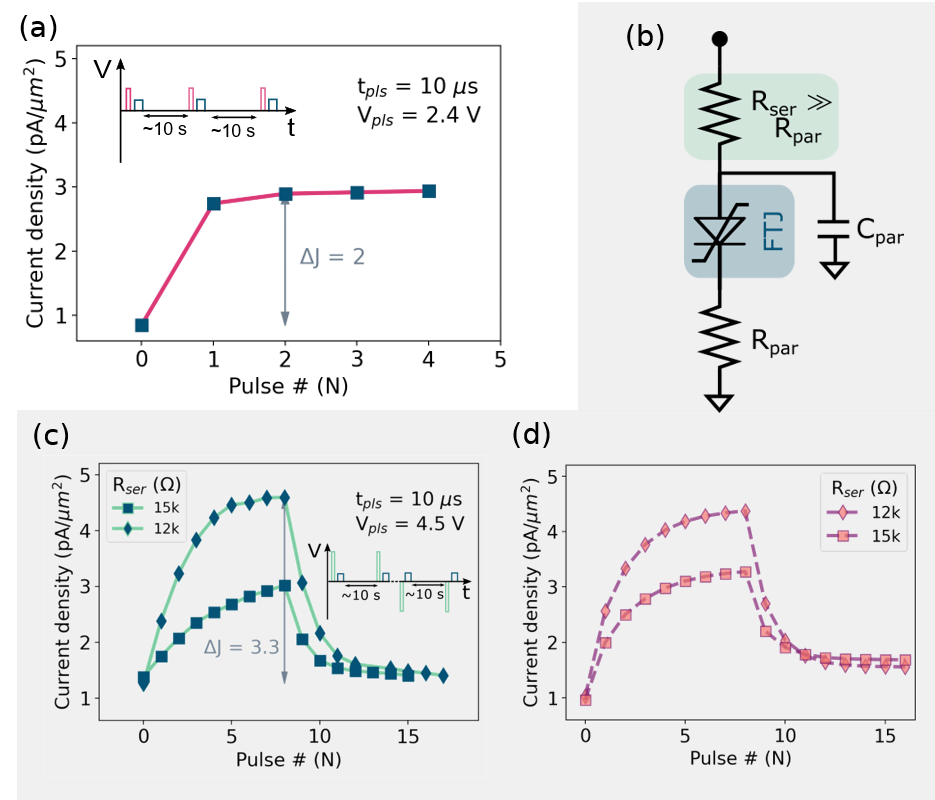}
	\captionsetup{font=footnotesize}
     \caption{(a) The weight update of an FTJ-1nm device when applying identical pulses with an arbitrary time delay between pulses. The switching parameters (shown inset) are 10\,$\mu$s width and 2.4\,V amplitude. (b) A schematic circuit diagram showing the insertion of a resistor in series with the FTJ. The resistor is inserted between the voltage source and the device. (c) The weight update of an FTJ device when applying identical pulses with an arbitrary time delay between pulses, for the circuit shown in (b). The pulse parameters (shown inset) are 10\,$\mu$s width and 4.5\,V amplitude. The weight update is plotted for a series resistance of 15\,k$\Omega$ (squares) and 12\,k$\Omega$ (diamonds). Measurements were performed on FTJ-1nm. (d) Simulated FTJ weight update behaviour using the FTJ VerilogA model calibrated on FTJ-1nm data, using the RC circuit sketched in (b).}
\label{Fig4_IdenticalResistor}
\end{figure}

 When applying an ideal square switching pulse to this 'RC' circuit, the external voltage will first drop over the resistor and the capacitor needs to charge to reach the respective coercive voltage $V_C$ before polarization switching can occur \cite{si2019ultrafast}. In addition, especially in the case of an FTJ a background leakage current $I_{leak}$ will always flow through the capacitor. Thus, switching will dominate on the pulse plateau, when $\frac{dV_{app}}{dt} = 0$ and with the series resistance chosen such that:
\begin{equation}
\frac{V_{app}-V_{max}}{I_{DE}(V_{max}) + I_{leak}(V_{max})+ I_{switch, max}} < R_{ser} \lesssim \frac{V_{app}-V_{min}}{I_{DE}(V_{min}) + I_{leak}(V_{min})}.
\label{Resistancevalue}
\end{equation}
Here, $V_{min}$ and $V_{max}$ denote the lower and the upper boundary of the coercive voltage distribution, respectively. Thus, within these limits, $R_{ser}$ will control the rate at which the polarization switches. Note, a higher value of $R_{ser}$ will greater restrict the ferroelectric switching, which could be compensated on the other hand by applying a higher voltage $V_{app}$ from outside. If $R_{ser}$ is too large, the capacitor cannot charge and polarization switching will not occur. It should be noted that such weight update schemes have been successfully implemented in resistive random-access memory (RRAM) systems \cite{moon2017improved}.

The above discussed concept was experimentally verified. For comparison, figure \ref{Fig4_IdenticalResistor}a shows the behaviour of an FTJ-1nm device when identical pulses are applied without connecting an external resistor. The pulse parameters are 2.4\,V, 10\,$\mu$s. The FTJ state is read out after each pulse, for 5 pulses. It can be seen that a large weight update ($\sim 2 pA/\mu m^2$, or $\sim$ 43$\%$ of the MW) occurs on the first pulse. Note that in contrast to the experiments shown in the previous section, the time delay between pulses is on the order of seconds, when no accumulative effects due to a residual field in the capacitor should be present \cite{siannas2023electronic}. Thus, under this condition the FTJ-1nm is operated with fully independent switching pulses where no cumulative weight update occurs. All domains with a coercive voltage below the applied voltage will switch, and subsequent similar pulses barely change the state, with a strongly diminishing effect from the second pulse. This behaviour can be described by the Preisach model \cite{bartic2001preisach}.

In contrast, in figure \ref{Fig4_IdenticalResistor}c, full switching pulses (4.5\,V, 10\,$\mu$s) are applied to an FTJ-1nm device, with resistors of different values connected in series before the biased electrode, according to the scheme shown in figure \ref{Fig4_IdenticalResistor}b. Clearly, a more gradual weight update can be observed even thought the switching conditions are still in a Preisach-like or non-cumulative regime. This results from the fact that in this current-limiting approach, the effective voltage over the ferroelectric layer increases with subsequent pulses. Obviously, for a broader switching peak the range of the effectively applied voltage for switching domains with lowest ($V_{min}$) towards highest ($V_{max}$) coercive voltage increases. The extension of this voltage range $V_{max} - V_{min}$ affects the required voltage drop over $R_{ser}$ and thus, while approaching $V_{max}$ the switching current limit decreases.

Furthermore note, that an asymmetric weight update occurs in the devices. In negative polarity, the weight update is more abrupt but slowly saturating, due to the lower Vc and broader switching distribution in negative polarities. Then, as the resistor value is increased, the weight update becomes more linear but the total polarization which can be reached decreases. This can be attributed again to an RC effect, whereby the voltage drop over the ferroelectric is limited by the large resistance and the capacitance of the setup (including device and parasitics), and does not reach a large enough value to switch all domains within the time width of the pulse. 

In the example of the FTJ used here, the read current is very small compared to the switching current ($I_{read} \ll I_{switch, max}$), so the series resistance does not interfere with the read operation for any value of R$_{ser}$. 

\subsection{Strategy towards linear weight update}
In order to achieve efficient, accurate training in a neural network, a synaptic device should linearly change its conductance state during its weight update \cite{ielmini2019emerging}. As seen in the previous example, the series resistor method, while successful in achieving gradual weight update with identical pulses, may lead to either a non-linear behaviour \textit{or} a significantly reduced MW. Furthermore, resistor values need to be specifically chosen depending on the device size, device capacitance and desired weight update properties. Finally, a general issue of this series-resistor based current limit is the fact that the controlled switching current decreases for domains that exhibit a larger coercive voltage, due to the fact that the current limit is voltage-dependent. In order to improve this non-linearity, a voltage-independent current limit would be desirable. 

 It has been demonstrated \textit{via} current-limited PFM measurements on BiFeO$_3$ that programming with an increasing current compliance allows multilevel switching \cite{lee2012multilevel}. In that work, the compliance was continuously increased in order to control the number of switching domains, and the scheme has recently been demonstrated on ferroelectric HfAlOx FeFETs \cite{kim2024unlocking}. 

 \begin{figure}[!hb]
\centering
		\includegraphics[width=\textwidth]{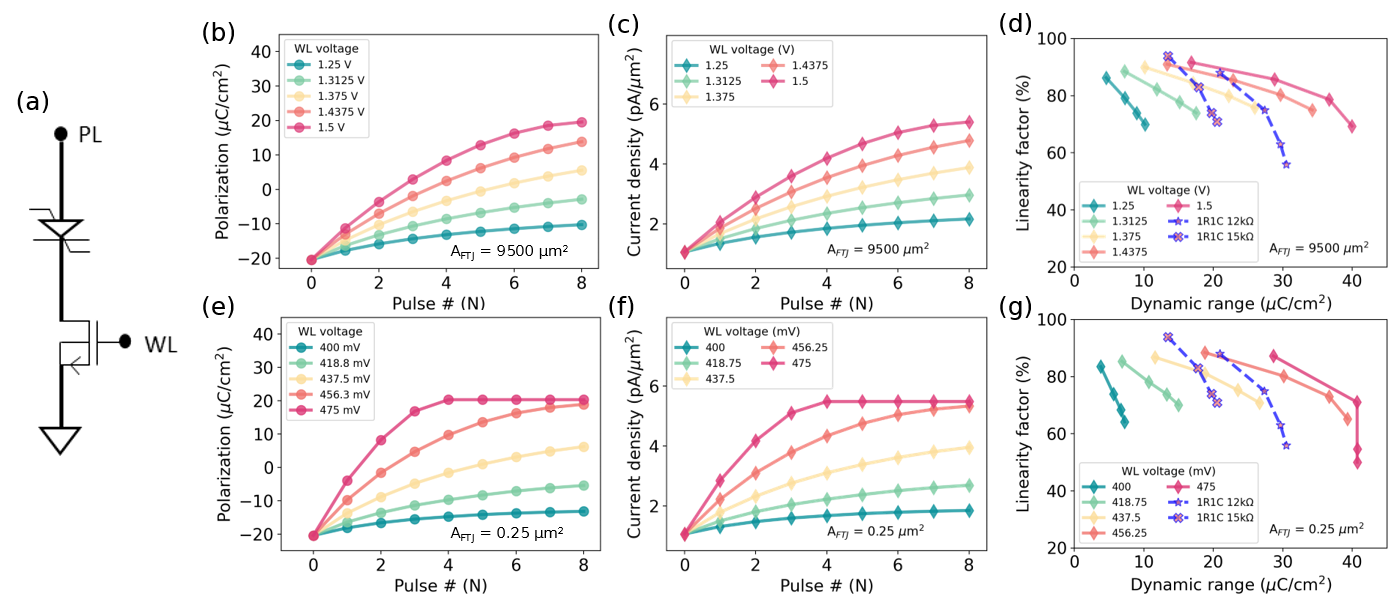}
	\captionsetup{font=footnotesize}
     \caption{(a) 1T1C circuit modelled for controlling the current during switching. Switching pulses of 4.5\,V, 10\,$\mu$s are applied on the plate line (PL) while a voltage applied on the word line (WL) limits the switching current. (b) Polarization and (c) read current density of an FTJ device simulated for 8 pulses at different WL voltages, for a 9500\,$\mu$m$^2$ device. (d) Linearity factor calculated using equation \ref{linearity} for different dynamic ranges (maximum switched polarization) for a 9500\,$\mu$m$^2$ device. The blue dashed lines show the extracted LF and dynamic range for the 1R1C data as a comparison. (e) Polarization and (f) read current density of an FTJ device simulated for 8 pulses at different WL voltages, for a 0.25\,$\mu$m$^2$ device. (e) LF for different dynamic ranges (maximum switched polarization) for a 0.25 \,$\mu$m$^2$ device. The blue dashed lines show the extracted LF and dynamic range for the 1R1C data on a 9500\,$\mu$m$^2$ device for reference.}
\label{Fig5_1T1C}
\end{figure}

Similarly, to increase the linearity of the weight update, we propose an improved solution, whereby the current reaching the ferroelectric capacitor is limited by the drain current of a transistor that is operated either in sub-threshold or saturation mode. In this case, the capacitor is connected between drain and the pulse voltage source (e.g. in a typical 1T1C FeRAM configuration), and the drain current is independent of the drain voltage. In contrast to the method demonstrated above, no changing current needs to be applied to the ferroelectric device: rather, the word line (WL) voltage of the transistor is kept constant, significantly simplifying the application of this scheme. 

In this case, experiments on large capacitors have not yet been enabled due to the difficulty of applying a current compliance in pulsed operation using commercial probe stations. Instead, the principle is demonstrated via Cadence simulations of simple bitcells. The ferroelectric device is modelled using a Preisach approach \cite{reuben2023low, gibertini2022ferroelectric}. 

In order to prove this concept, we demonstrate \textit{via} circuit simulations that multilevel behaviour can be achieved with identical pulses by applying a current compliance on-chip through an access transistor. In order to calibrate our simulation deck, in a first step the experimental results utilizing a series resistor as discussed in the previous section have been repeated in simulation. The FTJ model was calibrated to exactly replicate the PV-hystereis curve of the FTJ-1nm device under test (see figure S1). Then the same circuit as shown in figure \ref{Resistancevalue}b was implemented. Thereby, an additional parasitic cable capacitance of 500 pF was connected towards ground in order to replicate the cable capacitance from our experimental setup.
\ref{Fig4_IdenticalResistor}d reveals a good agreement between measurement and simulation. In this simulation identical pulses with an amplitude of 4.5\,V and a width of 10\,$\mu$s have been applied. In between each switching pulse a smaller read voltage pulse was utilized to monitor the FTJ read current. The small differences in the weight update behavior between experiment and simulation can be attributed to small deviations in the exact PV-curve between simulation and experiment, and the occurrence of potential imprint effects that so far are not included in our model \cite{vecchi2024evaluation, barbot2023dynamics}. 

In order to demonstrate improved weight update linearity, the same calibrated FTJ model was then implemented into a 1T1C circuit (figure \ref{Fig5_1T1C}a). Similar to in the case above, identical switching pulses have then been applied. This simulation was performed for varying gate voltages, thus resulting in different drain current limits. The resulting weight update behavior is depicted in figures \ref{Fig5_1T1C}b-d for a large device as measured above, and in figures \ref{Fig5_1T1C}e-g for a scaled device with nm side lengths. It should be noted that the physical mechanism of ferroelectric switching is field-driven and remains so in the 1R1C and 1T1C configurations. By applying a current compliance the device self-limits the voltage which develops across the ferroelectric layer, so that on each consecutive pulse, a higher voltage is applied across the ferroelectric device. These increasing voltages have been output from simulations and are plotted in figure S2. Extrinsic effects which modify the local electric fields of some portions of the ferroelectric, including charge trapping, residual fields and ionic motion in the electrodes may still be present, but the voltage across the ferroelectric will adjust while the amount of switched charge should remain more robust to extrinsic effects. Moreover, by controlling the current compliance, in a 1T1C cell the switched charge per pulse can be tuned through V$_{WL}$, thereby controlling the number of levels and the linearity.

In figures \ref{Fig5_1T1C}d,g, a linearity factor and a dynamic range (i.e. the range of total switched polarization) were extracted at different pulse numbers for the data in figures \ref{Fig5_1T1C}b,e. The linearity factors (LF) describe the linearity of the weight update behaviour and are calculated by: 
\begin{equation}
    LF = \frac{J_n - J_0}{J_{n/2} - J_0}/2
    \label{linearity}
\end{equation}
where J$_n$ represents the current density on the \textit{nth} switching pulse and J$_0$ represents the initial device current density i.e. in the fully \textit{Off} state. The parameter LF should be 1 for a linearly switching device and can be expressed as a percentage. According to figure \ref{Fig1_SwitchingBasic}, the dynamic range in $\mu$C/cm$^2$ is proportional to the FTJ MW. For a 1T1C circuit, LF reaches $\sim$ 86 \% at a dynamic range of 30 $\mu$C/cm$^2$ and falls to $\sim$ 69 \% at a dynamic range of 40 $\mu$C/cm$^2$. LF generally falls with increasing dynamic range, which happens more rapidly in scaled devices (figure \ref{Fig5_1T1C}g). In 1R1C operation, this reduction in linearity with increasing dynamic range is the most pronounced, due to the RC behaviour of the circuit. For a 15\,k$\Omega$ resistor, a LF of 71 \% is reached with a dynamic range of 20 $\mu$C/cm$^2$, while for a 12\,k$\Omega$ resistor, a dynamic range of 30 $\mu$C/cm$^2$ can be reached but with a reduced LF of 63 \%, indicating the trade-off present in this operation mode.

\begin{figure}[!hb]
\centering
		\includegraphics[width=\textwidth]{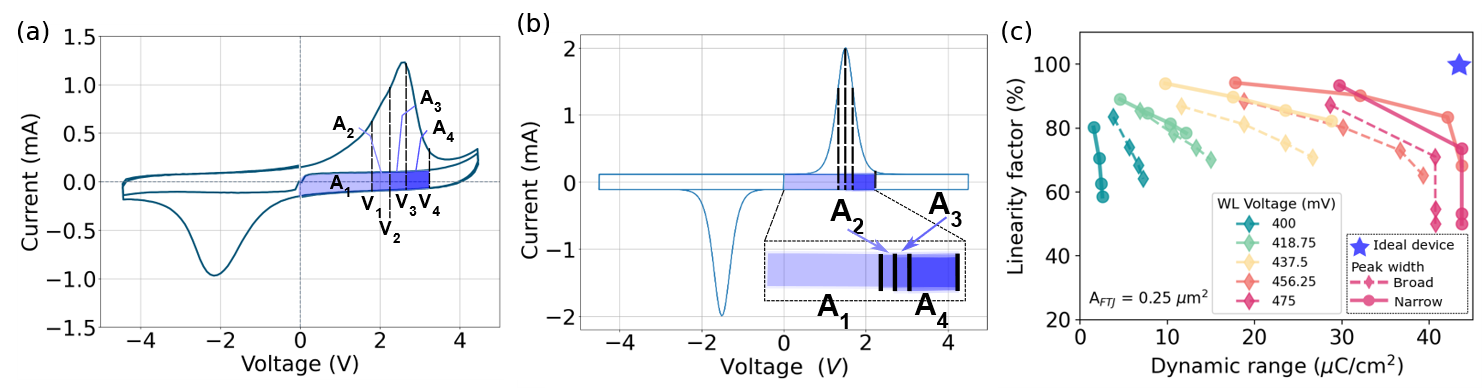}
	\captionsetup{font=footnotesize}
     \caption{(a) Area of the capacitive displacement current (proportional to overall charge displacement ) at different voltages for the measured I-V curves. (b) Area of the capacitive displacement current at different voltages for a simulated I-V curve with narrower switching peak. (c) LF and dynamic range extracted from the weight update behaviour of a 1T1C cell where the ferroelectric has a broad (dotted lines) or narrow (solid lines) switching peak.}
\label{Fig6_NarrowBroad}
\end{figure}

The remaining non-linearity can be attributed to the large distribution of switching voltages of the individual domains within the device. Even though the current limit is now independent on the voltage drop at the transistors drain terminal, the required charging of the dielectric background capacitance of the FTJ consumes an increasing amount of charge with increasing effective switching voltage, as shown in fig \ref{Fig6_NarrowBroad}a. At different voltages V$_n$, the area of the capacitive displacement current A$_n$ vary widely, which also leads to variations in the remaining current for domain switching. Hence, in order to obtain perfect linearization of the weight update, a very narrow ferroelectric switching peak is desirable. With a simulated narrow switching peak, the differences in areas A$_n$ are decreased, as illustrated in fig \ref{Fig6_NarrowBroad}b. Performing the same analysis of the linearity as discussed previously, it can be seen in \ref{Fig6_NarrowBroad}c that narrowing the switching peak leads to an expected increase in both dynamic range and LF at higher WL voltages. Hence, a dynamic range \textgreater 30 $\mu$C/cm$^2$ can be covered while maintaining a linearity \textgreater 90 \%. Finally, the strongly non-linear leakage current in FTJ devices can be detrimental to the linearity for a similar reason: with a changing applied voltage, the contribution of the current to leakage will increase. Therefore, leakage contributions should be minimized with respect to the displacement current, which can be achieved for example \textit{via} modifying the pulse frequency \cite{massarotto2022versatile}, or for some applications by operating the device preferentially in the reverse polarity. Using higher-frequency pulses should also allow to fit more states within the same dynamic range of switched polarization.

\section{Conclusions}
In conclusion, we have presented here several methods for achieving weight update in devices based on ferroelectric capacitors, with an FTJ device as a model example. Since the weight update occurs in the ferroelectric film, these methods are applicable (with possible circuit modifications) in other device types, including FeCAP and FeFET. 

To achieve weight update with non-identical pulses, a method was presented to find an automata (or state machine) model with 4 states, which could be extended for a higher number of states. Here, deterministic current levels could be defined in an FTJ device, and the required pulses to transition between them. These require complex circuitry to apply on-chip, where time and/or voltage of the applied pulses should be dynamically modified \cite{narayanan2022120db}. 

Two methods for achieving weight update with identical pulses and arbitrary time delays were also presented. The first, demonstrated experimentally, involves adding a resistor in series with the ferroelectric device. Depending on the resistor value, a high linearity could be achieved, but owing to an RC effect, higher resistor values limit the dynamic range or MW. Nonetheless, several distinct levels could be reached. This strongly differs from the case of identical pulses and short fixed time delays between pulses, where if the delay time between pulses is increased, the weight update diminishes \cite{siannas2023electronic, vecchi2024evaluation}. The time delay between pulses is again on the order of seconds, which makes the pulse sequence highly applicable for a number of neuromorphic computing elements \cite{ielmini2019emerging} including integrate-and-fire neurons \cite{gibertini2022ferroelectric} or as a synaptic element.

The final method, demonstrated \textit{via} Cadence simulations, involves applying an external current compliance. Here, the switching is limited by the drain current of a transistor operated either in sub-threshold or saturation mode. We have demonstrated that this works for different device sizes and simulations show this can achieve linearity up to 93\% covering most of the memory window. It is worth noting that this 1T1C method exhibits a trade-off between dynamic range and linearity, which can be reduced by careful selection of the WL voltage (i.e. the current compliance) and switching pulse parameters, and by engineering of the switching peak.

\section{Experimental section}
\subsection{Sample preparation} FTJ capacitors were fabricated by sputter deposition, atomic layer deposition (ALD), metal evaporation and dry etching (inductively coupled plasma, ICP). Sputtered metallic electrodes were employed consisting of W (30\,nm)/TiN (5\,nm)/TiAl (7.5\,nm)/TiN (5\,nm), forming TiAlN, on the BE and TiAlN on the TE. Al$_2$O$_3$ was deposited via ALD (precursor: TMA, oxidant: H$_2$O, T = 260\degree C) followed by 10.5\,nm HZO (precursors: HyALD, ZyALD, oxidant: O$_3$, T = 280\degree C). The samples were annealed at 450\degree C in Ar atmosphere for 30 mins. Capacitors were defined via the metal evaporation of Ti/Pt (10/25\,nm) which was used as a hard mask for rate-controlled ICP etching of the TE in BCl$_3$ chemistry. Chlorine chemistry is needed to etch the TiAlN electrodes but the rate must be measured since this chemistry can also etch HZO \cite{mauersberger2021single}. 

\subsection{Electrical measurements} All electrical analyses were performed on a Cascade Microtech probe station using a Keithley 4200 Parameter Analyser with Keithley 4225 RPMs which allow pulsed measurements down to a pulse width of 20\,ns. P-V loops were extracted using a PUND sequence at 20\,kHz. For multilevel experiments, pulse rise and fall times were chosen as 500\,ns, which is one order of magnitude larger than the resolution of the measurement equipment and should not be affected by any internal parasitics. Read-out pulses had a width of 50\,ms, which was chosen to further avoid influence of RC effects which delay the displacement current in large capacitors, obscuring the leakage current in the read-out. 

\subsection{Switching simulations} 
A VerilogA file modelling the ferroelectric behaviour with a Preisach approach was constructed (available upon reasonable request) and used in Cadence simulations. The FTJ parameters were fitted from experimental data and verified by comparison of simulated and measured P-V loops (figure S1). The modelled bitcells are shown schematically in figure \ref{Fig4_IdenticalResistor}b and \ref{Fig5_1T1C}a. These were constructed of the FTJ device model as well as a standard voltage source (analogLib), resistor (Primlib) and transistor (ne5). 


\medskip

\textbf{Acknowledgements} \par 
This work was supported by the DFG Priority Program ´Memristec' through the project ReLoFeMris (441909639), and by the European Commission through the project FIXIT (No. GA: 101135398). T.M. was financially supported out of the Saxonian State budget approved by the delegates of the Saxon State Parliament. Views and opinions expressed are those of the authors only and do not necessarily reflect those of the European Union.

\medskip

%
{\small
\bibliography{Manuscript}}


\end{document}


\maketitle 
\section{Simulated vs. measured switching curves}
\begin{figure}[!hb]
\centering
		\includegraphics[height=6cm]{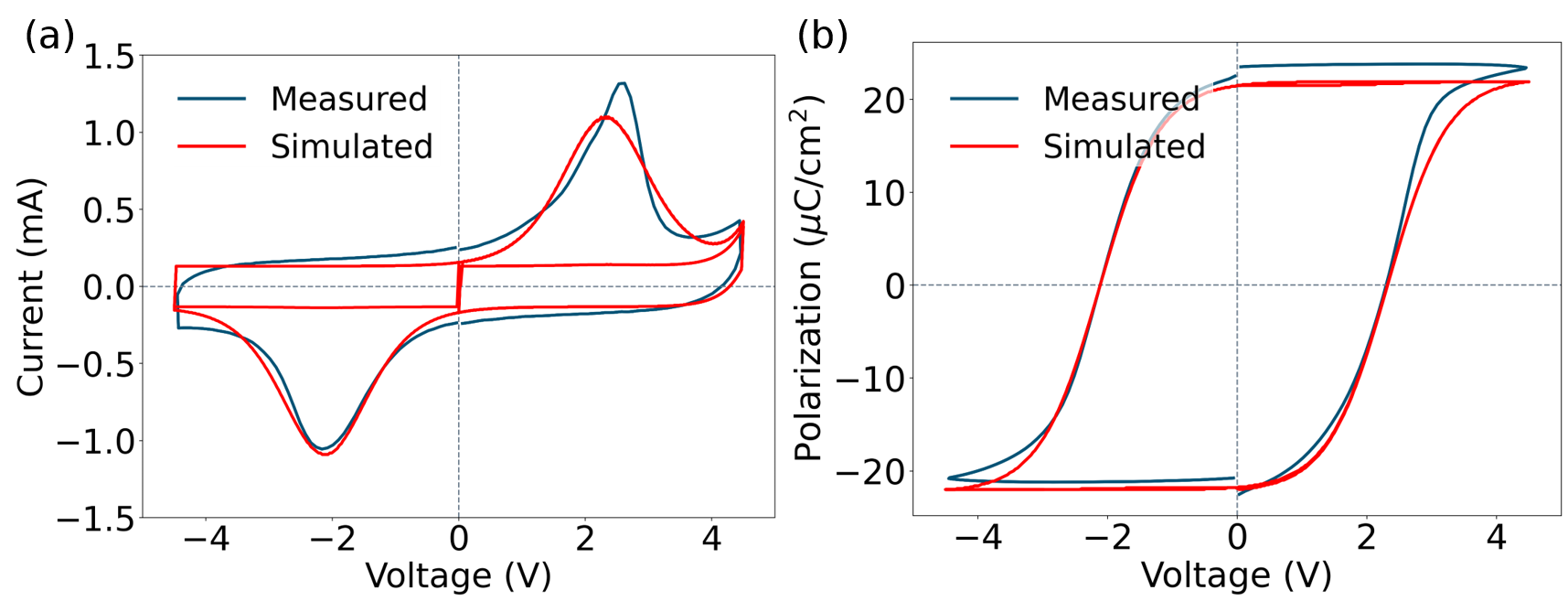}
	\captionsetup{font=footnotesize}
	\caption{(a) Current-voltage and (b) polarization-voltage curves for the 110\,nm diameter FTJ-1nm devices used in these experiments. Measurements were performed at 20\,kHz (blue lines). The simulated curves are plotted with red lines.}
\label{FigS1_PolSim}
\end{figure}

Figure \ref{FigS1_PolSim} shows the comparison between measured and simulated I-V (left) and P-V (right) curves for the FTJ-1nm device. The simulation was performed using the calibrated FTJ VerilogA model and captures the switching behaviour, including the increased leakage in positive polarity and the coercive voltages. The measured data show a narrower switching distribution in the positive polarity which is not fully captured by the model.
\newpage

\section{Voltage across the ferroelectric device}
\begin{figure}[!hb]
\centering
		\includegraphics[height=6cm]{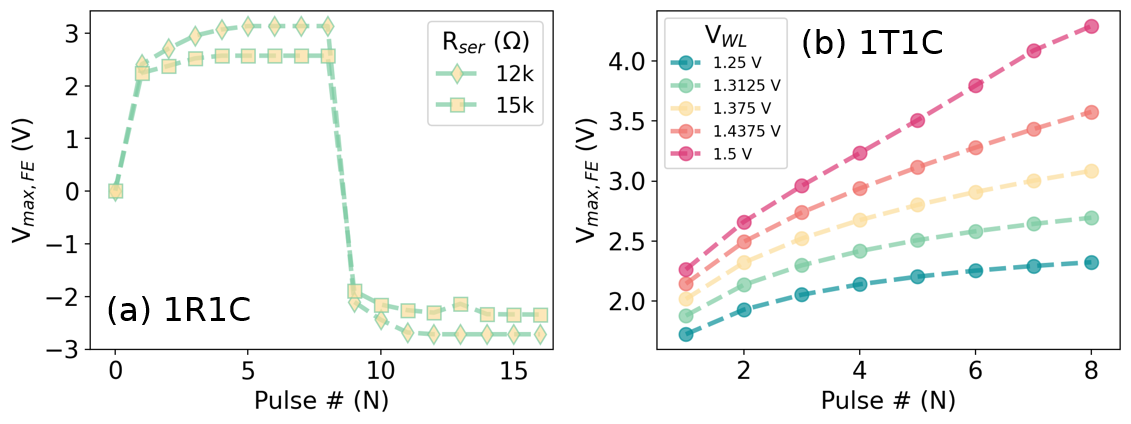}
	\captionsetup{font=footnotesize}
	\caption{Maximum voltage over the FTJ during each \textit{nth} switching pulse in the (a) 1R1C circuit under different series resistances and (b) 1T1C circuit under different WL voltages.}
\label{FigS2_Vmax}
\end{figure}

Figure \ref{FigS2_Vmax} shows the \textit{maximum} voltage over the ferroelectric during each switching pulse when programming under different conditions (varying R$_{ser}$/V$_{WL}$) in the 1R1C and 1T1C circuits, extracted from simulations. It should be noted that this V$_{max, FE}$ is only reached for a short time during each switching pulse, and the average applied voltage will be lower. Nonetheless, these figures demonstrate the self-limited increase in V$_{FE}$ on each subsequent pulse in both circuits. It should therefore be stressed that in a ferroelectric device, when applying a current compliance, the switching is still field-driven and the field across the ferroelectric develops depending on the applied current compliance. The broader range of V$_{max, FE}$ in the 1T1C circuit (V$_{WL}$ = 1.5\,V) indicates that a larger MW can be achieved.

\section{Calculation of linearity factors and dynamic range}
Linearity factors are extracted from the experimental data by the equation: 
\begin{equation}
    LF = \frac{J_n - J_0}{J_{n/2} - J_0}/2
    \label{linearity}
\end{equation}
where J$_n$ represents the current density on the \textit{nth} pulse. Between $n/2$ and $n$, the current density should increase by a factor of 2 in a linear system, leading to an ideal LF of 1.\\

The dynamic range reached with the \textit{nth} pulse is calculated by:
\begin{equation}
    range = P_n - P_0
    \label{range}
\end{equation}
where P$_n$ is the polarization state reached on the \textit{nth} pulse. The data presented here are plotted for n={2,4,6,8}. \\

It should be noted that for the 1R1C data, LF was calculated from the measured data while the dynamic range was extracted for the simulated data, again taking into account the proportionality between the dynamic range in terms of polarization charge and the MW in terms of FTJ current density.
\newpage